\documentclass[epj]{svjour}
\usepackage{graphics}
\usepackage{rotating,epic,eepic}
\begin{document}
\title{Recent Results from HEGRA}
\subtitle{
  Gamma-Ray Observations with the HEGRA Stereoscopic System of 5
  Cherenkov Telescopes}
\author{Niels G\"otting\inst{1}
  for the HEGRA Collaboration\inst{2}
}                     
\offprints{Niels G\"otting
  \\
  e-mail: Niels.Goetting@desy.de
  }          
\institute{
  Universtit\"at Hamburg, Institut f\"ur Experimentalphysik, 
  Luruper Chaussee 149, D-22761 Hamburg, Germany
  \and
  http://www-hegra.desy.de/hegra
  }
\date{Received: date / Revised version: date}
%
\abstract{
  The HEGRA collaboration has achieved outstanding results
during the operation of the six imaging atmospheric Cherenkov
telescopes from 1996 to 2002. The experimental work pioneered
the field of TeV $\gamma$-ray astronomy with observations
during partial moon time and mainly by applying the stereoscopic
observation mode using a system of five Cherenkov telescopes.
Concerning Galactic objects the HEGRA observations have led
to a precise measurement of the energy spectrum of the Crab
nebula between 0.5~and 80\,TeV, the detection of the first shell 
type supernova remnant in the Northern hemisphere (Cassiopeia A)
and the investigation of the yet unidentified HEGRA TeV
$\gamma$-ray source TeV J2032+4130 in the Cygnus region. In
addition, a large fraction of the Galactic plane has been studied
during dedicated scans.
Following the most precise measurements of the energy spectra of
the well known extragalactic objects Mkn 421 and Mkn 501, the
blazars H\,1426+428 and 1ES\,1959+650 have just been established as sources
of TeV photons in the last two years. Extensive multi-wavelength campaigns 
have been successfully performed and spectroscopy of these
four objects gives important clues for the understanding of the
nonthermal emission processes and also on the optical to 
infrared part of the spectrum of the extragalactic background light.
Recently, strong evidence for the nearby giant radio galaxy
M\,87 being a TeV $\gamma$-ray emitter has been obtained.
Some of these results are highlighted in this article.
} 
\maketitle
\section{Introduction}
\label{intro}
The HEGRA\footnote{HEGRA stands for {\em High Energy Gamma-Ray Astronomy}}
collaboration has operated six imaging atmospheric Cherenkov telescopes
(IACTs)
on the Canary island of La Palma (28.75$^\circ$\,N, 17.89$^\circ$\,W) at
a height of 2200\,m above sea level. The prototype telescope CT\,1
\cite{mirzoyan_1994} was used as a stand alone detector introducing for
the first time observations during partial moon time. With the operation of 
the 5 telescopes CT\,2 - CT\,6 in stereoscopic observation mode (HEGRA
IACT system) \cite{daum_1997} HEGRA has
pioneered the stereoscopic technique adopted by most of the next generation
experiments. The stereoscopic observation of an extended air shower, 
i.\,e.~the simultaneous measurement of the Cherenkov light initiated by the
particle cascade in the atmosphere with several telescopes under different
viewing angles, allows for an unambiguous reconstruction of the shower
direction, the impact point of the shower axis on the observation level
and the height of the shower maximum on an event by event basis. This leads
to an improved angular and energy resolution along with a significantly improved
$\gamma$/hadron separation. Furthermore, the coincidence method results
in a strong suppression of the background from night sky light 
and from local muons. The
stereoscopic observation mode in combination with the relatively large
field of view of the HEGRA telescopes also allows for a simultaneous 
observation of events from well defined background regions and for the 
performance of sky searches in the whole field of view 
(see e.\,g.~\cite{puehlhofer_scan_icrc}).
The sensitivity achieved with the HEGRA IACT system is a 10\,$\sigma$
detection within 1 hour for a source with a flux of 1 Crab. The operation of
the telescopes CT\,2~-~CT\,6 was terminated at the end of the year 2002.
%
%
%
\boldmath
\section{Galactic TeV $\gamma$-Ray Sources}
\unboldmath
\label{section_galactic}
%
%
\begin{figure}[t]
  %
  %
  \includegraphics[width=0.48\textwidth]{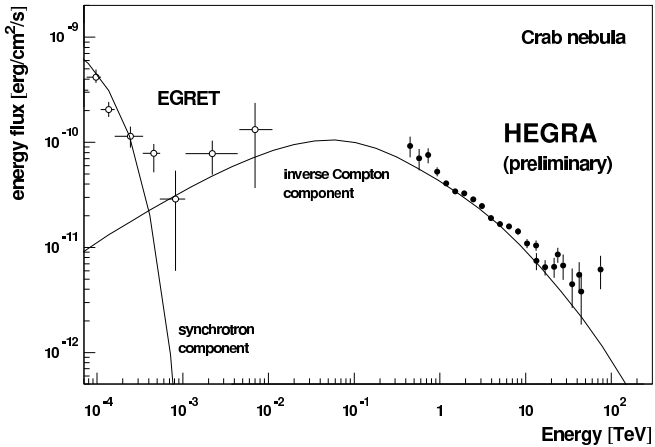}

\caption{
  The unpulsed energy spectrum of the Crab nebula ($E >$ 100\,MeV) as 
  measured by the EGRET experiment (open circles) as well as with
  the HEGRA IACT system (filled circles). A deep observation of nearly 400 hours
  with the HEGRA telescopes allowed for a measurement of the Crab spectrum in the
  energy range from 0.5~to 80\,TeV~\cite{hegra_crab_icrc}. The solid lines 
  indicate model expectations (from~\cite{hegra_crab_icrc}) for synchrotron and 
  inverse Compton radiation.
  }
\label{hegra_crab}       
\end{figure}
%
%
\subsection{Highest Photon Energies from the Crab Nebula}
The Crab nebula is the standard candle for TeV $\gamma$-ray astronomy in the
Northern hemisphere. Due to its outstanding role for calibration purpose of
the HEGRA IACT system a very long observation time of nearly 400 hours allowed
to measure the energy spectrum of the Crab nebula in the wide energy range 
from 0.5 up to 80\,TeV \cite{hegra_crab_icrc} (see Fig.~\ref{hegra_crab}). 
The detection of this object above 50\,TeV with a significance 
$ > 5\,\sigma$ in the HEGRA data makes it the $\gamma$-ray source detected at 
the highest photon energies so far.
%
%
\subsection{Detection of TeV J2032+4130 in Cygnus}
\begin{figure}[t!left panel]
  \includegraphics[width=0.48\textwidth]
    {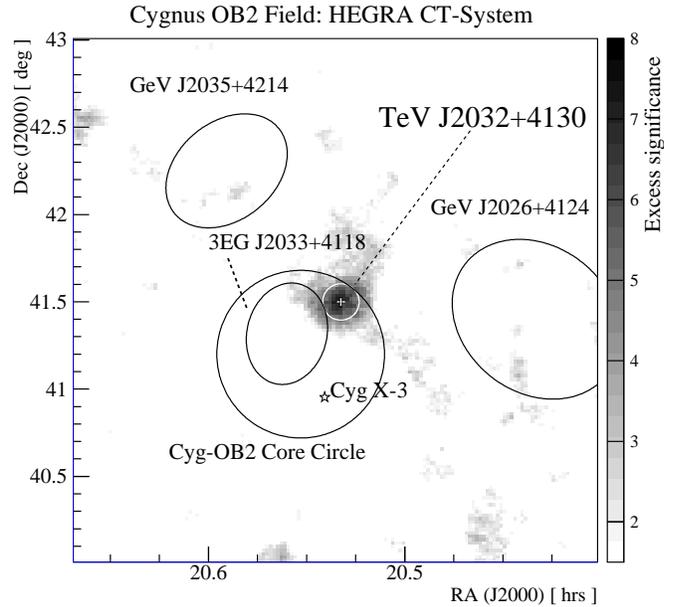}
  \caption{
    Skymap of the Cygnus region with the center of gravity of the TeV 
    $\gamma$-ray excess and the 2\,$\sigma$ error circle for TeV~J2032+4130 
    \cite{hegra_tevj2032_icrc}.
    Also marked are the 95\,\% error ellipses of three nearby EGRET GeV
    sources, the core of Cygnus OB2 and the location of Cygnus X-3.
    }
  \label{hegra_tev_source}
\end{figure}

Recently, with TeV J2032+4130 a yet unidentified (i.\,e.~no counterpart at
radio, optical nor X-ray energies) TeV $\gamma$-ray source has been detected
above 7\,$\sigma$ with the HEGRA IACT system in a 
direction about 0.5$^\circ$ to the North of Cygnus X-3 
\cite{hegra_tevj2032_icrc} (see Fig.~\ref{hegra_tev_source}). The 
object shows a hard spectrum (d$N$/d$E \propto E^{-1.9}$) and is possibly 
extended (on a 3\,$\sigma$ level). Several $\gamma$-ray production 
mechanisms are discussed and it may turn out that this first unidentified
source plays an important role in the search for the Galactic 
accelerators of the cosmic radiation. The HEGRA detection of TeV J2032+4130
points out that more up to now unexplored regions of the
nonthermal universe may be studied soon with the new generation of Cherenkov
telescopes.
%
%
\boldmath
\section{Extragalactic TeV $\gamma$-Ray Sources}
\unboldmath
\label{section_extragalactic}
%
%
\subsection{The TeV Blazars H\,1426+428 and 1ES\,1959+650}
H\,1426+428 is the most distant TeV $\gamma$-ray source established so far and
was detected by HEGRA (and also by the Whipple and CAT groups) 
in 1999/2000 and again in 2002 at a lower flux level
\cite{hegra_h1426_paper2}. Due to its large distance ($z = 0.129$) that is a 
factor of four
greater than the distances of the well known blazars Mkn 421 and Mkn 501 
a precise $\gamma$-ray spectrum of this third TeV blazar may be used to infer 
the density of the extragalactic background light (EBL) in the optical to 
near-infrared range due to the TeV photon absorbing pair production process 
$\gamma_{\mbox{\tiny TeV}} + \gamma_{\mbox{\tiny EBL}} 
\rightarrow e^+ + e^-$. Indeed, the present spectrum of H\,1426+428 as 
measured with the HEGRA IACT system already indicates a modulation of the
spectral shape \cite{hegra_h1426_paper2}.

The blazar 1ES\,1959+650, reported in 1999 by the 7\,TA collaboration to be a 
weak TeV $\gamma$-ray emitter (3.9\,$\sigma$), has been detected in the 
years 2000/2001 at 5.2\,$\sigma$ with the HEGRA IACT system in a deep 
exposure of 94 hours at a value of 5.3\,\% of the Crab flux. Subsequently,
strong outbursts ($> 20\,\sigma$) have been observed in May and 
July 2002 \cite{hegra_1es1959_paper} making 1ES\,1959+650 the fourth established TeV 
blazar with third best event statistics. The large HEGRA 1ES\,1959+650 data set
also allowed to determine the energy spectra during high ($> $ 1 Crab) and
low ($< $ 0.5 Crab) flux levels. Furthermore, the detection of the strong
flares in 2002 has initiated an extended multi-wavelength campaign combining
contemporaneously taken data from radio, optical, X-ray, and $\gamma$-ray 
telescopes. Using the light curves and spectral measurements over this broad
range of energies it is possible to study in detail the
nonthermal emission and particle acceleration processes inside the
relativistic jet of this object \cite{henric_1959_paper}.
%
%
\subsection{Detection of the Radio Galaxy M\,87}
\begin{figure*}[t]
  %
  %
  \includegraphics[width=0.48\textwidth]{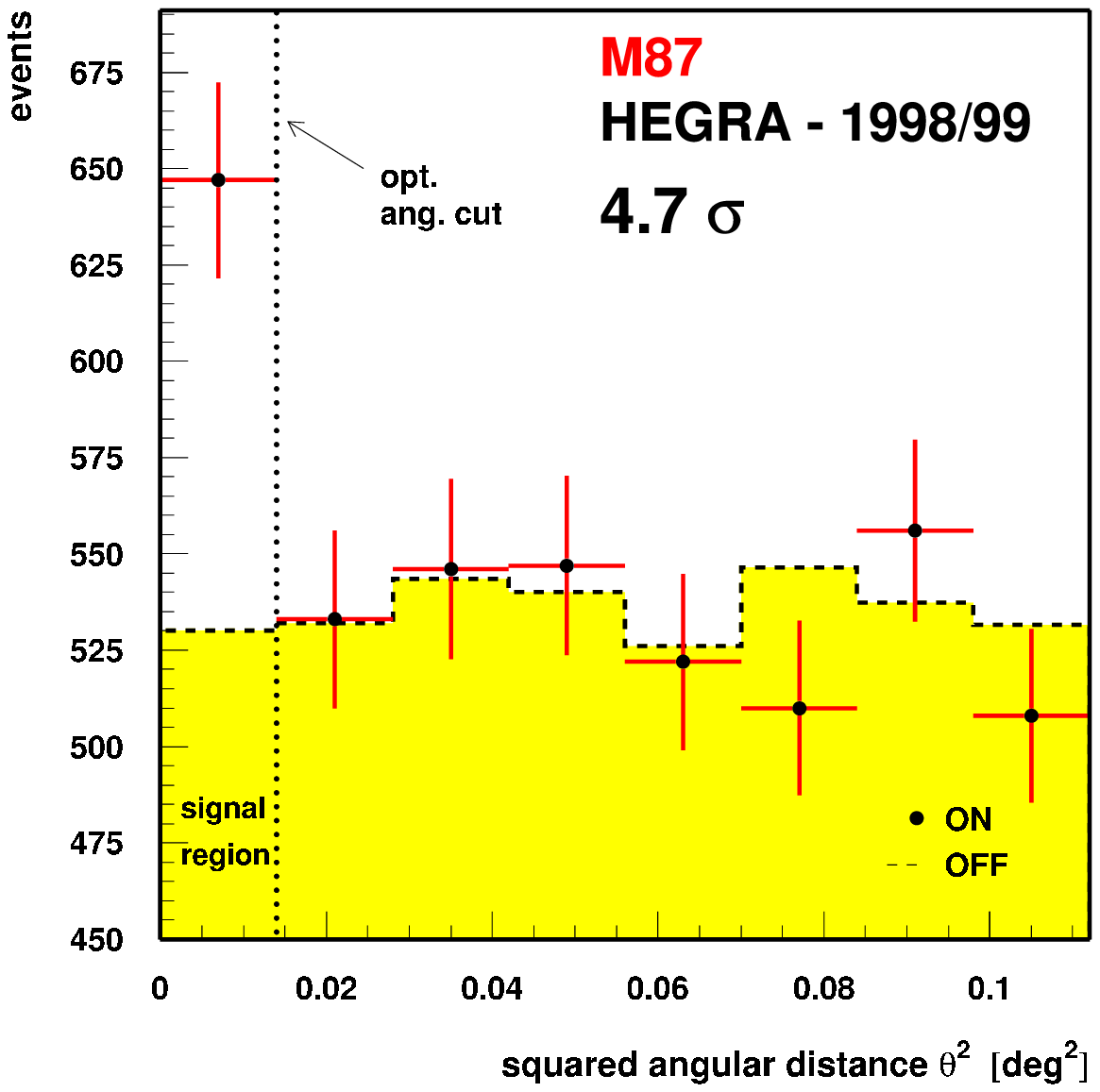}
  \hfill
  \includegraphics[bbllx=74, bblly=142, bburx=422, bbury=491, clip=, width=0.48\textwidth]{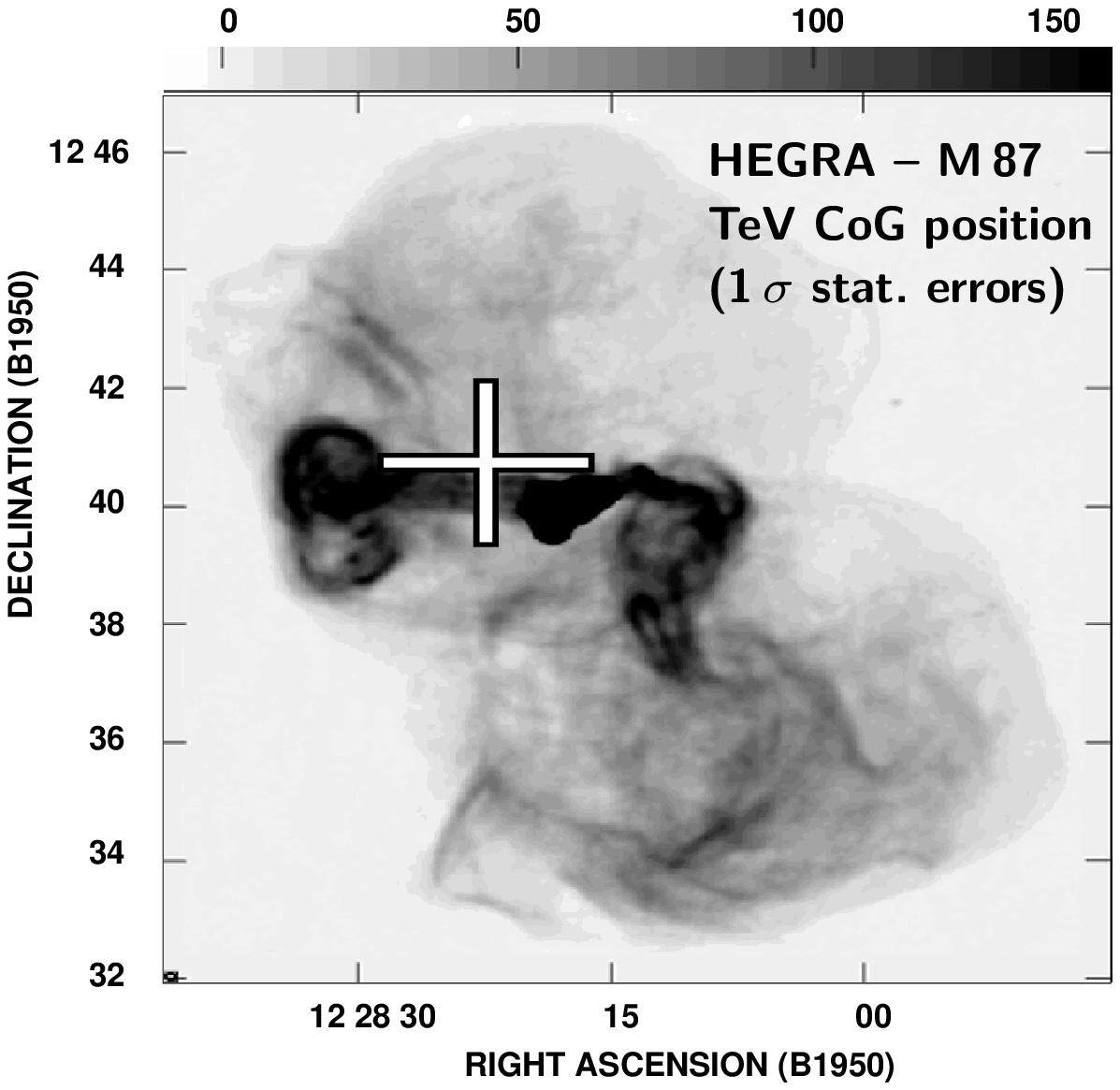}
\caption{
  Left panel: 
  Number of events vs.~squared angular distance $\Theta^2$ to the core
  position of M\,87 as observed in the years 1998 and 1999 with the HEGRA 
  IACT system. The dots show the ON-source events, while the histogram
  gives the background estimate.
  Indicated by the vertical dotted line is
  the optimum angular cut as determined from nearly contemporaneous Crab
  observations at similar zenith angles. The significance of the M\,87 excess 
  amounts to 4.7\,$\sigma$.
  Right panel: Radio image of M\,87 at 90\,cm showing the structure of the 
  M\,87 halo. The center of gravity position of the TeV $\gamma$-ray excess
  from the HEGRA M\,87 observations is marked by the cross 
  indicating the statistical 1\,$\sigma$ errors \cite{hegra_m87_icrc}.
  }
\label{hegra_m87_fig}       
\end{figure*}
The nearby giant radio galaxy M\,87 has been speculated for about 40 years now 
to be a powerful accelerator of cosmic rays including the highest energy 
particles observed in the universe (e.\,g.~\cite{ginzburg}). Furthermore, M\,87
is also considered as a source of TeV $\gamma$-rays from the hypothetical 
neutralino annihilation process.
At its center M\,87 contains a supermassive black hole of $2-3\times 10^9$~sun 
masses. The axis of the prominent relativistic kpc scale jet (also showing regions of 
superluminal motion well studied at radio, optical and X-ray frequencies) has 
an angle of 10-40$^\circ$ to the observer's line of sight 
in contrast to the blazars with their jets directly pointing to Earth. 

M\,87 has been observed with the HEGRA IACT system as one of the prime 
candidates for TeV $\gamma$-ray emission from the class of radio galaxies.
The deep HEGRA exposure of M\,87 of 77 hours in 1998/1999 has revealed for 
the first time a significant excess of photons above a mean 
energy threshold of 760\,GeV corresponding to (3.7 $\pm$ 0.8)\,\% of the Crab 
flux \cite{hegra_m87_paper,hegra_m87_icrc} (see Fig.~\ref{hegra_m87_fig}, 
left panel). 
Following recent improvements of the data analysis the signal detected with the
HEGRA telescopes has now a significance of 4.7\,$\sigma$.
On the basis of the limited event statistics 
the signal is compatible with a point-like source for the HEGRA IACT system. The
center of gravity (CoG) position of the HEGRA M\,87 TeV excess is shown in
Fig.~\ref{hegra_m87_fig}, right panel. Within the large statistical errors,
the CoG is consistent with the M\,87 core position, although a small shift of 
the source position cannot be ruled out.

M\,87 is the first Active Galactic Nucleus (AGN) beyond the
well-known blazar subclass being detected at TeV energies. With M\,87 the
AGN subclass of radio galaxies becomes an important new topic of TeV $\gamma$-ray
astronomy.
The TeV flux from M\,87 detected with the HEGRA IACT system can be
accomodated in different models, 
e.\,g.~within the Synchrotron Proton Blazar model
predicting protons to be accelerated to energies $\ge 10^{19}$\,eV
\cite{reimer_m87_spb_icrc}.
This makes M\,87 a promising target for the new generation of
Cherenkov telescope projects 
like the HEGRA succeeding H$\cdot$E$\cdot$S$\cdot$S telescope system
and the MAGIC telescope.

%
%
%
%
%
%
\begin{acknowledgement}
{
  The support of the German Federal Ministry for Research and Technology BMBF 
  and of the Spanish Research Council CICYT is gratefully acknowledged. 
  We thank the Instituto de Astrof\'{\i}sica de Canarias (IAC)
  for the use of the HEGRA site at the Observatorio del Roque de los 
  Muchachos (ORM) and for supplying excellent working conditions on La
  Palma.
}
  
\end{acknowledgement}
%
%

\end{document}